\documentclass[reprint,superscriptaddress,amsmath,amssymb,aps,notitlepage,floatfix,longbibliography]{revtex4-2}

\usepackage{graphicx}
\usepackage{color}
\usepackage{comment}

\usepackage{dcolumn}
\graphicspath{{./Figures_ASHE}}
\usepackage{upgreek}
\usepackage{gensymb}
\usepackage{float}
\usepackage{bm}
\usepackage{xcolor}
\usepackage{physics}
\usepackage{multirow}
\usepackage{array}

\begin{document}

\title{Spin current generation by acousto-electric evanescent wave}

\author{Takuya Kawada}
\email[]{takuyakawada@g.ecc.u-tokyo.ac.jp}
\affiliation{Department of Physics, The University of Tokyo, Bunkyo, Tokyo 113-0033, Japan}
\affiliation{Department of Basic Science, The University of Tokyo, Meguro, Tokyo 152-8902, Japan}

\author{Masashi Kawaguchi}
\affiliation{Department of Physics, The University of Tokyo, Bunkyo, Tokyo 113-0033, Japan}

\author{Kei Yamamoto}
\affiliation{Advanced Science Research Center, Japan Atomic Energy Agency, Tokai, Ibaraki 319-1195, Japan}

\author{Hiroki Matsumoto}
\affiliation{Department of Physics, The University of Tokyo, Bunkyo, Tokyo 113-0033, Japan}
\affiliation{Institute for Chemical Research, Kyoto University, Uji, Kyoto 611-0011, Japan}

\author{Ryusuke Hisatomi}
\affiliation{Institute for Chemical Research, Kyoto University, Uji, Kyoto 611-0011, Japan}
\affiliation{Center for Spintronics Research Network (CSRN), Kyoto University, Uji, Kyoto 611-0011, Japan}
\affiliation{PRESTO, Japan Science and Technology Agency, Kawaguchi, Saitama 322-0012, Japan}

\author{Hiroshi Kohno}
\affiliation{Department of Physics, Nagoya University, Chikusa, Nagoya 464-8602, Japan}

\author{Sadamichi Maekawa}
\affiliation{Advanced Science Research Center, Japan Atomic Energy Agency, Tokai, Ibaraki 319-1195, Japan}
\affiliation{RIKEN Center for Emergent Matter Science, Wako, Saitama 351-0198, Japan}
\affiliation{Kavli Institute for Theoretical Sciences, University of Chinese Academy of Sciences, Beijing 100049, China}

\author{Masamitsu Hayashi}
\email[]{hayashi@phys.s.u-tokyo.ac.jp}
\affiliation{Department of Physics, The University of Tokyo, Bunkyo, Tokyo 113-0033, Japan}
\affiliation{Trans-Scale Quantum Science Institute (TSQS), The University of Tokyo, Bunkyo, Tokyo 113-0033, Japan}

\newif\iffigure
\figurefalse
\figuretrue

\date{\today}
	
\begin{abstract}
	We experimentally demonstrate that a spin current can be induced by the acousto-electric evanescent wave, an electric field associated with surface acoustic waves (SAWs) that decay along the surface normal. A previous study showed that a magnetic-field-dependent dc voltage (acoustic voltage) emerges in heavy metal (HM)/ferromagnet (FM) bilayers under excitation of SAWs. The effect, referred to as the acoustic spin Hall effect, was understood by assuming a SAW-induced ac spin current rectified by the oscillation of the FM layer magnetization and the inverse spin Hall effect. However, the mechanism of the spin current generation remained unidentified.
	Here we measure the acoustic voltage as a function of the SAW propagation direction relative to the crystalline orientation of a LiNbO$_3$ substrate. We find that the magnetic field angle dependence of the acoustic voltage exhibits a phase shift depending on the SAW propagation direction. The result is consistently explained in terms of the acousto-electric evanescent wave generating the spin current in HM layer via the spin Hall effect, thus clarifies the origin of the acoustic spin Hall effect.  
\end{abstract}

\maketitle

\begin{figure}[t]
	\centering
	\includegraphics[scale=0.075]{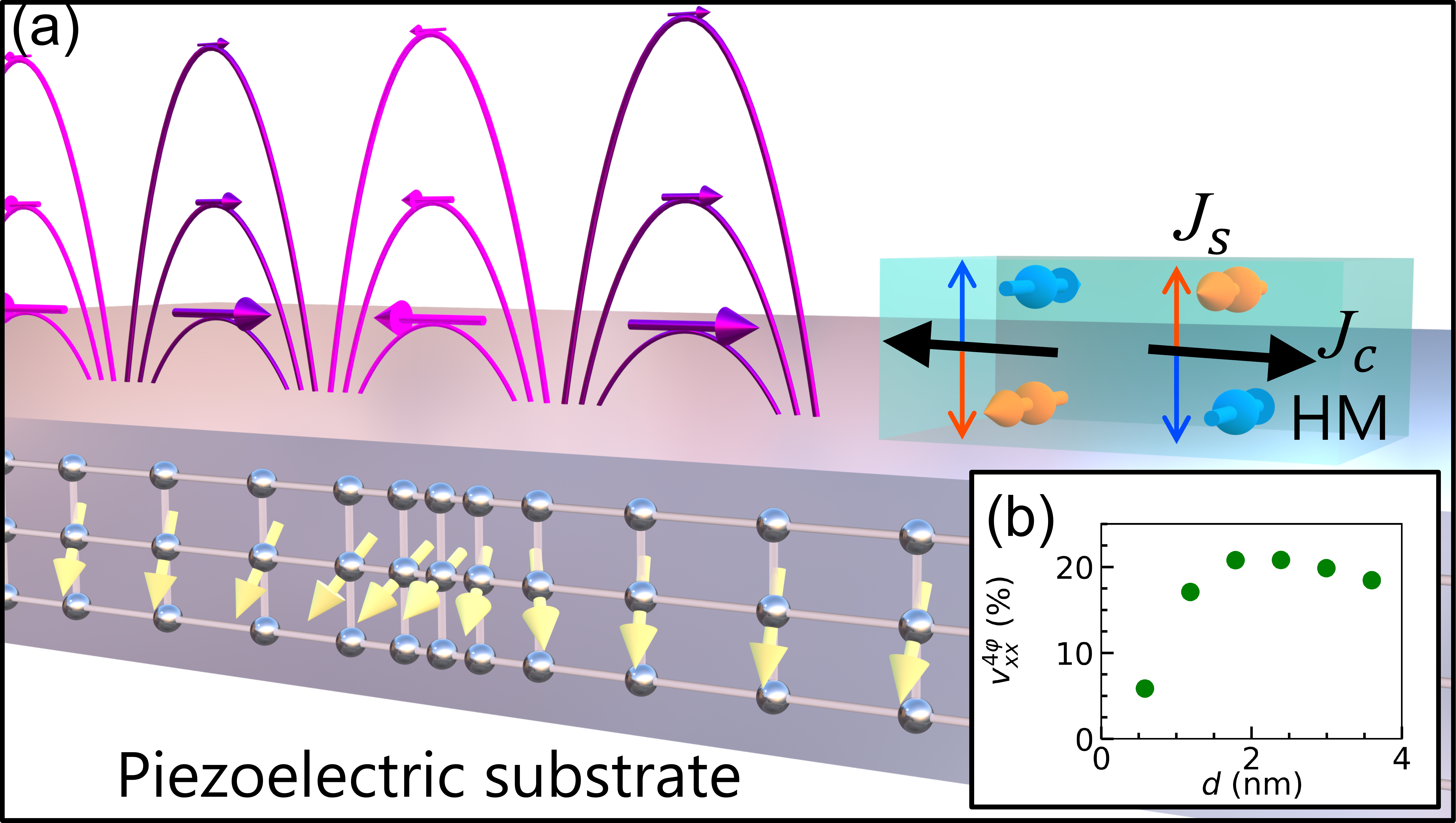}
	\caption{
		(a) A schematic illustration of the acousto-electric evanescent wave that travels with a SAW on the surface of a piezoelectric substrate and the spin current generated by the field in a heavy metal (HM) thin film.
		The grey spheres and the light yellow arrows show atoms and their local spontaneous polarization, respectively, of the substrate.
		The purple lined arrows represent the electric force associated with the acousto-electric evanescent wave.
		The light blue box shows the film deposited on the substrate. The blue and orange spheres represent electrons with opposite spins flowing in opposite direction in the film induced by the field and the spin Hall effect.
		(b) W thickness ($d$) dependence of the normalized acoustic voltage $v_{xx}^{4\varphi}\equiv\Delta V_{xx}^{4\varphi}/V_{xx}^{0}$ measured for a W($d$)/CoFeB bilayer.
		Reproduced from Ref.~\cite{kawada2021sciadv}.
		\label{fig:main_Fig1}
	}
\end{figure}

Spin mechatronics~\cite{goennenwein2014ssc,matsuo2017jpsj} is an emerging field that investigates the interaction between spin angular momentum and mechanical motions. Surface acoustic waves (SAWs), localized sound modes near solid surfaces, are frequently employed to generate rf mechanical vibrations. SAWs can be coherently excited by applying an rf signal to an interdigital transducer (IDT) fabricated on a piezoelectric substrate. Previous studies have demonstrated that SAWs can couple with electron spins via magnetoelastic coupling~\cite{weiler2011prl,dreher2012prb,sasaki2017prb,kuss2020prl,hatanaka2022prapp,matsumoto2024acs,hwang2024prl,chumak2010prb,lynos2023prl,cao2024ncomm,mi2024apl,nii2024arXiv,chen2023ncomm,yang2024ncomm,foerster2017ncomm,casals2020prl} and spin-vorticity coupling~\cite{matsuo2013prb,kobayashi2017prl,kurimune2020prl,mingxian2023}, leading to spin wave excitation and non-equilibrium spin accumulation. 
A recent experimental study showed that SAWs induce an ac spin current in metallic thin films with significant spin-orbit interaction~\cite{kawada2021sciadv}. In this effect, termed as the acoustic spin Hall effect, the SAW-induced ac spin current was found to possess the following properties (see the light blue box of Fig.~\ref{fig:main_Fig1}(a)): (a) The spin current flows parallel to the surface normal with its polarization orthogonal to both the flow and SAW propagation directions; (b) The effect is observed in heavy metals but not in materials with small spin-orbit interaction, such as Cu; (c) The spin current is distributed uniformly in the film thickness direction. If the electric field from the piezoelectric substrate is efficiently screened by conduction electrons and decays rapidly in conductors, the last property suggests that the spin current have a mechanical origin rather than an electromagnetic one~\cite{funato2021jmmm,mahfouzi2022prl}. Based on this observation, it was proposed that the source of the SAW-induced spin current in heavy metals is proportional to the time derivative of the lattice displacement~\cite{kawada2021sciadv}. However, the microscopic mechanism underlying the spin current generation has remained unidentified.

In this work, we demonstrate that the origin of the acoustic spin Hall effect is an acousto-electric evanescent wave, an electric field that is localized near the substrate surface and propagates at the speed of the SAW (see Fig.~\ref{fig:main_Fig1}(a)).
As reported previously, the ac spin current manifests itself in a magnetic-field-dependent dc voltage, referred to as the acoustic voltage hereafter, in heavy metal (HM)/ferromagnet (FM) bilayers~\cite{kawada2021sciadv}.
The acoustic voltage arises from the magnetoresistance-mediated rectification of the ac spin current with an in-phase oscillation of the FM layer magnetization via SAW strain and magnetoelastic coupling. 
The experimental setup employed in the previous study, however, was unable to identify the root cause of the SAW-induced spin current, whether it is mechanical or electrical.
Here we measure the acoustic voltage as a function of the SAW propagation direction with respect to the crystalline principal axis of the piezeoelectric substrate.
We find that the magnetic field angle dependence of the acoustic voltage is phase-shifted as the SAW propagation direction is varied, implying the electrical origin of the acoustic spin Hall effect. 
The phase shift is reproduced if we assume that the ac spin current is proportional to the in-plane electric field $E_x^\mathrm{TM}$ associated with the acousto-electric evanescent wave. 

First, we outline the strategy to identify the origin of the spin current. 
When a SAW traverses a HM/FM bilayer, the acoustic voltage emerges via a process that resembles the spin Hall magnetoresistance (SMR)~\cite{Nakayama2013SMR}.
Suppose the SAW generates an ac spin current in HM that is uniform in the film normal direction.
The spin current diffuses to the HM/FM interface and gets reflected.
The degree of reflection depends on the FM layer magnetization direction: it is the largest when the electron spin polarization of the incoming spin current is parallel to the magnetization while it takes a minimum when they are orthogonal to each other.
The SAW also changes the magnetization direction via its strain and the magnetoelasticity of the FM layer.
If the magnetization direction oscillates in phase with the SAW-induced ac spin current, the latter is rectified upon reflection at the interface, thus causing a dc spin current that flows away from the interface.
The reflected dc spin current is converted to a dc electric current via the inverse spin Hall effect~\cite{saitoh2006} of the HM layer, which is detected as the dc acoustic voltage. 

\begin{figure}[t]
	\centering
	\begin{minipage}{1.0\hsize}
		\centering
		\includegraphics[scale=0.075]{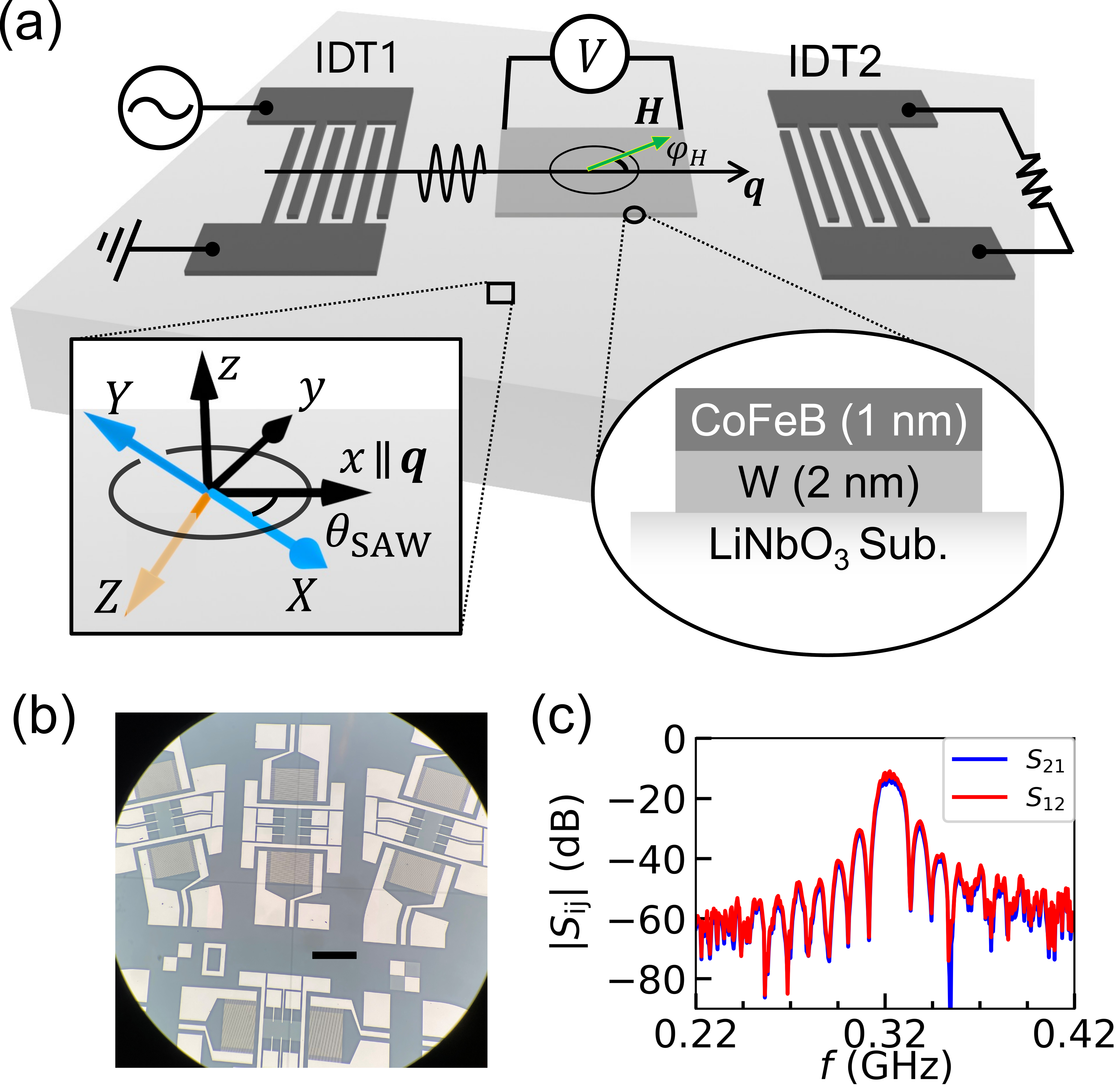}
	\end{minipage}
	\caption{
		(a) A schematic representation of the measurement setup. The black and green arrows denote the propagation direction of the SAW ($\bm{q}$) and the in-plane magnetic field ($\bm{H}$), respectively.
		The left inset shows the crystalline axes $XYZ$ of LiNbO$_3$ substrate ($Z$ corresponds to the spontaneous polarization) and the lab coordinate system ($xyz$). Here we set $x$ along $\bm{q}$. 
		The film stacking of the HM/FM bilayer employed in this study is shown in the right inset.
		(b) An optical microscopy image of representative devices. The black scale bar in the image corresponds to a length of $500\ \upmu$m. The bright areas show the electrodes and the IDTs, whereas the dark grey regions indicate the substrate. The light grey-rectangles located in the center of the IDTs represent the HM/FM thin films.
		(c) A typical SAW transmission spectrum. $\theta_\mathrm{SAW}$ is set to $0^\circ$. The blue (red) curve shows the transmission spectrum when the SAW propagates from IDT1 to IDT2 (IDT2 to IDT1).
	}
	\label{fig:main_Fig2}
\end{figure}

The uniformity of the spin current across the film thickness was inferred from the HM thickness dependence of the acoustic voltage.
Figure~\ref{fig:main_Fig1}(b) presents an example from the previous study~\cite{kawada2021sciadv}. The acoustic voltage reaches its maximum near the spin diffusion length of the HM layer, which is consistent with the drift-diffusion model assuming a uniform spin current~\cite{chen2013prb}. However, it was not possible to determine the mechanism of spin current generation, whether it has a mechanical or electrical origin. 
To overcome this issue, we study the acoustic voltage as a function of the angle ($\theta_\mathrm{SAW}$) between the SAW propagation direction $\bm{q}$ and the crystalline $X$-axis of the LiNbO$_3$ piezoelectric substrate (see inset to Fig.~\ref{fig:main_Fig2}(a)).
Due to the crystalline anisotropy of the piezoelectric tensor, it is known that the relative phase between the SAW's mechanical properties (e.g., lattice displacement, strain, vorticity) and electrical properties (e.g., electric field, electric displacement field) shows a strong $\theta_\mathrm{SAW}$ dependence~\cite{datta1986book}.
In contrast, the relative phase among the mechanical properties or the electrical properties hardly depends on $\theta_\mathrm{SAW}$.
As described above, the acoustic voltage arises only when the ac spin current varies in phase with the strain-induced magnetization oscillation (note that the latter is a mechanical property).
If the spin current originates from one of the SAW electrical properties, the acoustic voltage will show a significant $\theta_\mathrm{SAW}$ dependence, whereas it will hardly depend on $\theta_\mathrm{SAW}$ if the source has a mechanical origin.
Thus, the $\theta_\mathrm{SAW}$ dependence of the acoustic voltage, together with model calculations, will allow us to determine the mechanism of the SAW-induced spin current.

The measurement setup and the coordinate system are schematically shown in Fig.~\ref{fig:main_Fig2}(a).
A HM/FM bilayer film is grown on a Y$+128^\circ$-cut LiNbO$_3$ substrate using rf magnetron sputtering.
The film structure is sub./W(2)/CoFeB(1)/MgO(2)/Ta(1) (thickness in units of nanometers), where the top MgO/Ta layer is used as the capping layer to prevent oxidation of the film.
Hereafter, the film is referred to as W/CoFeB.
Standard optical lithography and Ar ion etching are employed to pattern a rectangle element from the film.
Contact electrodes and IDTs made of conducting metals are formed using a liftoff process (see Supplemental Material~\cite{kawada2024ashesuppl_arXiv}, Sec.~I~A for the detail).

\begin{figure}[htb]
	\centering
	\begin{minipage}{1.0\hsize}
		\centering
		\includegraphics[scale=0.15]{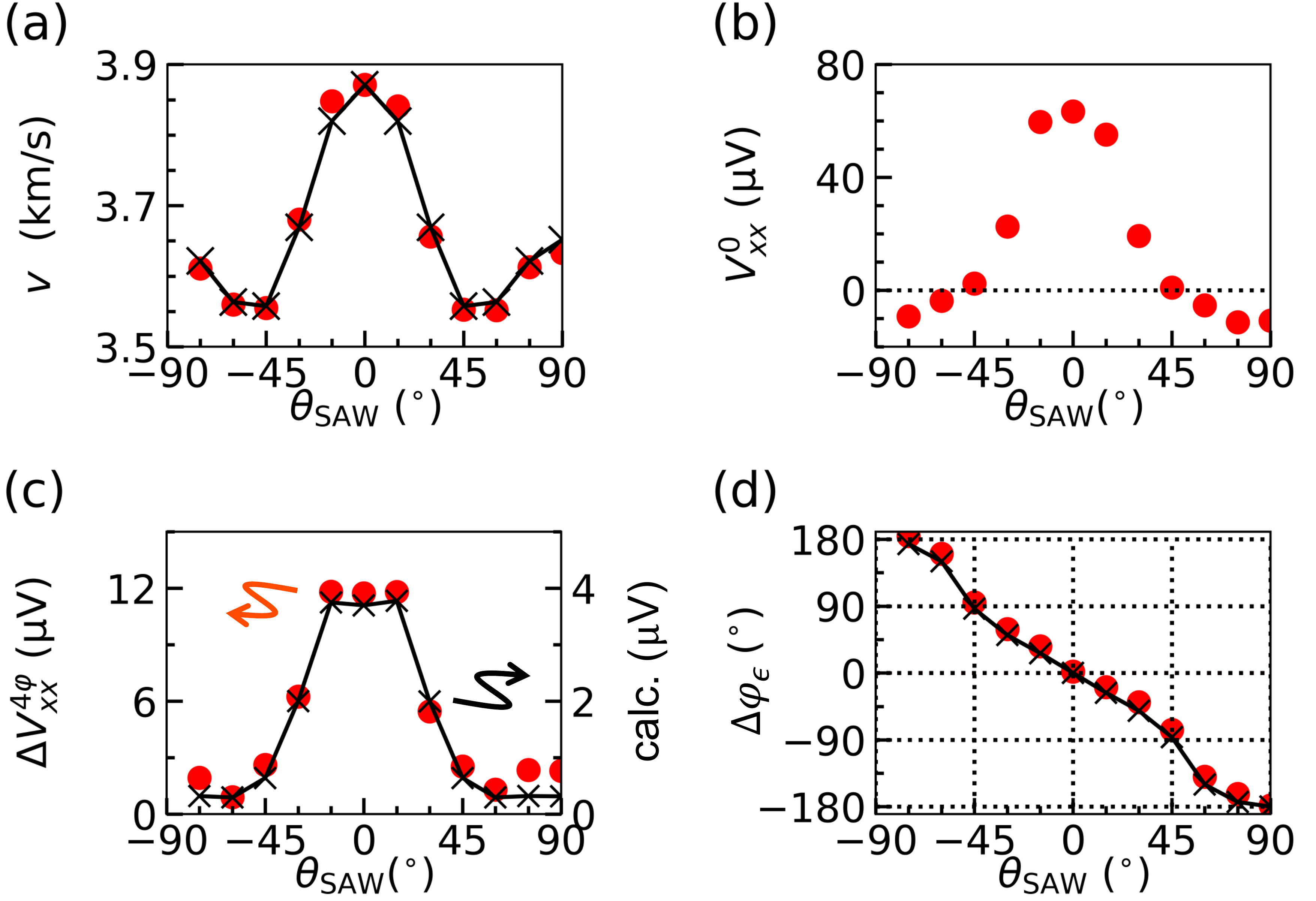}
	\end{minipage}
	
	\caption{
		The $\theta_\mathrm{SAW}$ dependence of $v$ (a), $V_{xx}^{0}$ (b), $\Delta V_{xx}^{4\varphi}$ (c), and $\Delta \varphi_\epsilon$ (d) for W/CoFeB.
		$V_{xx}^{0}$, $V_{xx}^{4\varphi}$, and $\Delta \varphi_\epsilon$ are obtained from the fit to the data using Eq.~(\ref{eq:fit_4theta}).
		The data are obtained from measurements using a magnetic field of 60 mT.
		The rf signal is applied to IDT1 with frequency at the SAW resonance and power of 13 dBm.
		Black crosses in (a,c,d) show results of the model calculation.
	}
	\label{fig:main_Fig4}
\end{figure}

An optical microscope image of the device is shown in Fig.~\ref{fig:main_Fig2}(b).
A number of SAW devices with identical geometry are fabricated on the same substrate.
The orientation of each device is varied with respect to the LiNbO$_3$ crystalline $X$-axis to change the SAW propagation direction $\bm{q}$ in the substrate plane.
The SAW transmission spectra are measured using a vector network analyzer (VNA) to obtain the SAW resonance frequency.
A typical spectrum is shown in Fig.~\ref{fig:main_Fig2}(c).
All devices show a clear transmission peak corresponding to the fundamental SAW mode.
The SAW resonance frequency $f_c$ is estimated by a Lorentzian fitting.
The SAW velocity $v$ is then derived from the dispersion relation $v=f_c \lambda$, where $\lambda$ is the SAW wavelength defined by the pitch of the IDT wires and is fixed to $12\ \upmu$m.
The $\theta_\mathrm{SAW}$ dependence of $v$ is shown in Fig.~\ref{fig:main_Fig4}(a), which is consistent with the velocity of the Rayleigh-type SAW obtained in previous studies~\cite{weser2020ultrason}.

The longitudinal and transverse acoustic voltages, denoted by $V_{xx}$ and $V_{yx}$, respectively, are measured as a function of the magnetic field angle $\varphi_H$ and magnitude $H$ while an rf signal is applied to IDT1 from a signal generator (see Fig.~\ref{fig:main_Fig2}(a)).
The average value of the rf power is fixed to 13 dBm and its frequency is set to $f_c$.
The measurement is performed under two different magnetic field magnitudes of approximately 60 mT and 18 mT.
Details of the measurements are described in Sec.~I~C of the Supplemental Material~\cite{kawada2024ashesuppl_arXiv}. 

The $\varphi_H$ dependence of $V_{xx}$ is shown in Fig.~\ref{fig:main_Fig3} (see Supplemental Material~\cite{kawada2024ashesuppl_arXiv}, Sec.~II~C  for the results of $V_{yx}$). $V_{xx}$ shows a sinusoidal change whose period is $90^\circ$ in all cases.
The amplitude of the sinusoidal modulation is smaller when the magnitude of the magnetic field is larger.
This is caused by the reduction in the strain-induced magnetization oscillation amplitude when the magnetic field is increased~\cite{kawada2021sciadv}. 
Interestingly, the $\varphi_H$ dependence of the acoustic voltage varies with $\theta_\mathrm{SAW}$.
For instance, $V_{xx}$ changes in the form of $\sin^2 2\varphi_H$ for $\theta_\mathrm{SAW}=0^\circ$ whereas it varies as $\sin 4\varphi_H$ for $\theta_\mathrm{SAW}=45^\circ$.

Guided by the above observation, we fit the data using the following function:
\begin{equation}
	\begin{aligned}
		V_{xx} = V_{xx}^{0} + \Delta V_{xx}^{4\varphi} \sin 2\varphi_H \sin \qty(2 \varphi_H - \Delta \varphi_\epsilon).
		\label{eq:fit_4theta}
	\end{aligned}
\end{equation}
$\Delta V_{xx}^{4\varphi}$ represents the amplitude of the $90^\circ$-periodic component.
$\Delta \varphi_\epsilon$ is a phase shift to account for the change in the $\varphi_H$ dependence with $\theta_\mathrm{SAW}$.
\begin{figure}[t]
	\centering
	\begin{minipage}{1.0\hsize}
		\centering
		\includegraphics[scale=0.15]{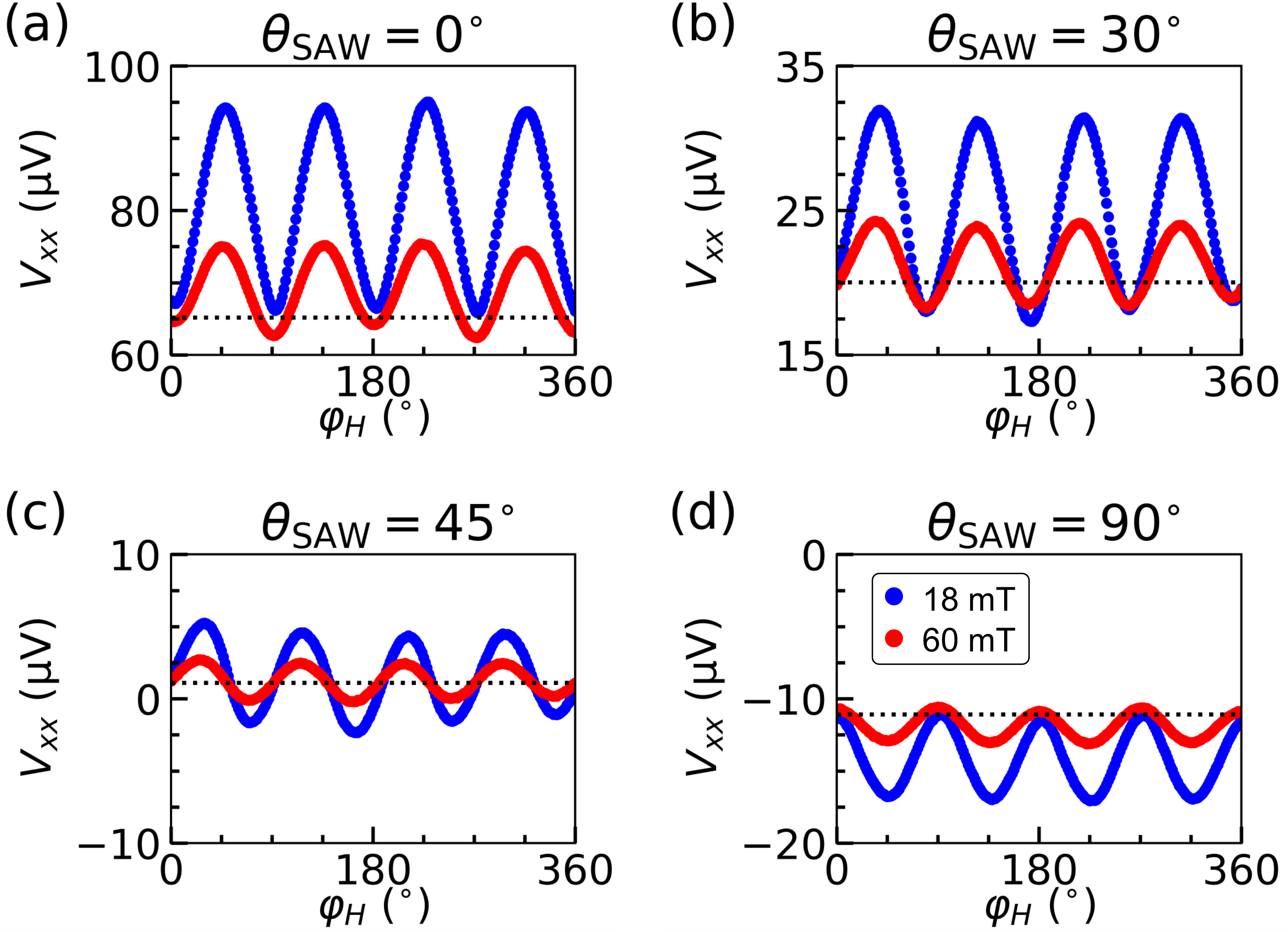}
	\end{minipage}
	
	\caption{
		(a-d) The magnetic field angle ($\varphi_H$) dependence of the acoustic voltage $V_{xx}$ for W/CoFeB.
		The red (blue) dots show data when the magnitude of the magnetic field is set to 60 mT (18 mT). $\theta_\mathrm{SAW}$ is denoted at the top of each panel.
		The black dotted lines indicate the field-independent component $V_{xx}^{0}$ defined by Eq.~(\ref{eq:fit_4theta}).
		The rf power is fixed at 13 dBm, and the frequency is set to the SAW resonance frequency of 323 MHz (a), 305 MHz (b), 297 MHz (c), and 302 MHz (d).
	}
	\label{fig:main_Fig3}
\end{figure}
$V_{xx}^{0}$, $\Delta V_{xx}^{4\varphi}$, and $\Delta \varphi_\epsilon$, obtained from the fitting to the $\varphi_H$ dependence of $V_{xx}$ with Eq.~(\ref{eq:fit_4theta}), are plotted against $\theta_\mathrm{SAW}$ in Fig.~\ref{fig:main_Fig4}(b-d).
All the parameters show significant dependences on $\theta_\mathrm{SAW}$.
First, $V_{xx}^{0}$, indicated by the horizontal dotted lines in Fig.~\ref{fig:main_Fig3}, changes its sign at a certain angle between $\abs{\theta_\mathrm{SAW}}=45^\circ$ and $60^\circ$.
It was previously reported that $V_{xx}^{0}$ originates from the acousto-electric current, the dc charge flow generated by SAW~\cite{weinreich1957pr,ingebrigtsen1970jap,rotter1998apl,miseikis2012apl,nichols2024jap}.
Conventional theory suggests that the sign of the acousto-electric current is the same regardless of $\theta_\mathrm{SAW}$ \cite{ingebrigtsen1970jap}.
Hence the sign reversal of $V_{xx}^{0}$ observed in this work cannot be explained with such theory.
As the focus of this work is on $\Delta V_{xx}^{4\varphi}$ and $\Delta \varphi_\epsilon$, we address the $\theta _\mathrm{SAW}$ dependence of $V_{xx}^{0}$ in a different occasion.
$\Delta V_{xx}^{4\varphi}$ and $\Delta \varphi_\epsilon$ also show significant dependences on $\theta_\mathrm{SAW}$. $\Delta V_{xx}^{4\varphi}$ is roughly symmetric with respect to $\theta_\mathrm{SAW}=0^\circ$, whereas $\Delta \varphi_\epsilon$ monotonically and continuously changes from $-180^\circ$ to $+180^\circ$ when $\theta_\mathrm{SAW}$ is varied from $+90^\circ$ to $-90^\circ$.

The results presented in Fig.~\ref{fig:main_Fig4}(c,d) establish that the acoustic voltage varies significantly with $\theta_\mathrm{SAW}$, suggesting that the spin current is of electrical origin.
To account for the $\theta_\mathrm{SAW}$ dependence of the acoustic voltage quantitatively, we extend the model developed previously~\cite{kawada2021sciadv} to include the effects of all the strain components associated with the SAW.
The following expression is obtained for the acoustic voltage (see Supplemental Material~\cite{kawada2024ashesuppl_arXiv}, Sec.~III for the details of the derivation): 
\begin{equation}
	\begin{gathered}
		\Delta V_{xx}^{4\varphi} = L \rho_\mathrm{N}  r_{xx}^{2\varphi} J_{\epsilon} \frac{b}{\mu_0 H M_\mathrm{S}},\\
		J_{\epsilon} \equiv A_\mathrm{s} \sqrt{\qty(\frac{1}{2}\mathrm{Re}\qty[\qty(X_\mathrm{s})^{*} \cdot \epsilon_{xx}])^2+\qty(\frac{1}{2}\mathrm{Re}\qty[\qty(X_\mathrm{s})^{*}\cdot 2\epsilon_{xy}])^2},
		\label{eq:J1e_amp_rev}
	\end{gathered}
\end{equation}
where $\rho_\mathrm{N}$ is the resistivity of the HM layer, $r_{xx}^{2\varphi}$ is the spin Hall magnetoresistance ratio of the HM/FM bilayer, $L$ is the length of the film along the SAW propagation, $b$ and $M_\mathrm{S}$ are the magnetoelastic coupling energy density and the saturation magnetization of the FM layer, respectively, and $\mu_0$ is the vacuum permeability. $\epsilon_{xx}$ and $\epsilon_{xy}$ are the longitudinal and shear strain induced by the SAW.
$X_\mathrm{s}$ denotes the hypothetical source of spin current such as a component of electric field, and $A_\mathrm{s}$ is a prefactor for $X_\mathrm{s}$.
The phase shift $\Delta \varphi_\epsilon$ is given as
\begin{equation}
	\begin{gathered}
		\cos \Delta \varphi_\epsilon \equiv \frac{A_\mathrm{s}}{2J_{\epsilon }}\mathrm{Re}\qty[\qty(X_\mathrm{s})^{*} \cdot \epsilon_{xx}] ,\\
		\sin \Delta \varphi_\epsilon \equiv \frac{A_\mathrm{s}}{2J_{\epsilon }}\mathrm{Re}\qty[\qty(X_\mathrm{s})^{*}\cdot 2\epsilon_{xy}] .
		\label{eq:J1e_phase-shift_rev}
	\end{gathered}
\end{equation}
See Supplemental Material~\cite{kawada2024ashesuppl_arXiv}, Eq.~(S35) for the explicit form of $\Delta \varphi_\epsilon$ using the relative phase between $X_\mathrm{s}$ and $\epsilon_{xx}, \epsilon_{xy}$.
Our task is now to identify $X_\mathrm{s}$ that reproduces the experimental results shown in Fig.~\ref{fig:main_Fig4}(c,d).

We perform a model calculation of electrical properties of the SAW (as well as mechanical properties for confirmation) and the resulting acoustic voltage~\cite{campbell1968ieee,ingebrigtsen1969jap,ingebrigtsen1970jap}: see Supplemental Material~\cite{kawada2024ashesuppl_arXiv}, Sec.~III, for the details of calculation. The calculated SAW velocity, plotted in Fig.~\ref{fig:main_Fig4}(a) with black crosses, reproduces the experimental results, thus justifying the model.
We find that all the experimental results can be accounted for if we take $X_\mathrm{s}=E_x^\mathrm{TM}$, the electric field along $x$ originating from the acosuto-electric evanescent wave with decay length $\sim \lambda$. Since the film thickness is much smaller than $\lambda$, $E_x^\mathrm{TM}$ is almost uniform along $z$.
Note that $E_x^{\rm TM}$ is not electrostatically screened by the conduction electrons since it is an electromagnetic transverse field~\cite{kawada2024arXiv}.
The calculated $\Delta \varphi _\epsilon$, shown by the black crosses in Fig.~\ref{fig:main_Fig4}(d), agrees well with the experimental data.
We check whether other choices for $X_\mathrm{s}$ may reproduce the data, and find that only $E_x^\mathrm{TM}$ can account for the experimental results: see Supplemental Material~\cite{kawada2024ashesuppl_arXiv}, Fig.~S5. 
Note that the time derivative of the lattice displacement vector, proposed as the source of the SAW-induced spin current previously~\cite{kawada2021sciadv}, can account for the experimental results only when $\theta_\mathrm{SAW}=0^\circ$.

The amplitude of the acoustic voltage $\Delta V_{xx}^{4\varphi}$ is also estimated by the model calculation. 
First, we find $|E_x^\mathrm{TM}|=4.1$ V/cm and $|\epsilon_{xx}|=7.5 \times 10^{-5}$ for the SAW excited by the rf power of 18 mW along $\theta_\mathrm{SAW}=0^\circ$ when the substrate surface is covered with a film whose resistivity and thickness are $150\ \upmu\Omega\cdot$cm and $3$ nm, respectively.
$\Delta V_{xx}^{4\varphi}$ is then calculated from Eq.~(\ref{eq:J1e_amp_rev}) with the electrical conductivity $ \sigma _c =\rho_\mathrm{N}^{-1}$ substituted for $A_\mathrm{s}$, using the following parameters: $L=460\ \upmu$m, $\rho_\mathrm{N}=150\ \upmu\Omega\cdot$cm~\cite{kawada2021sciadv}, $\mu_0 H=60$ mT, $b=-4.7$ MJ/m$^3$~\cite{matsumoto2024acs}, and $M_\mathrm{S}=1.3$ MA/m~\cite{kawada2023apl}.
$r_{xx}^{2\varphi}$ is obtained from the present experiments: we find $\sim$0.9 \% for all devices with different $\theta_\mathrm{SAW}$ (see also Fig.~S2(b) of Supplemental Material~\cite{kawada2024ashesuppl_arXiv}). 
The calculated $\Delta V_{xx}^{4\varphi}$ is shown by the black crosses in Fig.~\ref{fig:main_Fig4}(c), reproducing the $\theta _\mathrm{SAW}$ dependence well while the amplitude is off roughly by a factor of three.
The discrepancy is likely due to the use of literature values of material parameters (both the LiNbO$_3$ substrate and the thin films) in the calculation. Given that there are no free parameters to adjust, we consider this to be a satisfactory agreement.
We thus conclude that the origin of the spin current is the ac electric field that accompanies the SAW and the spin Hall effect of the HM layer.
We remark that the acousto-electric evanescent wave is in fact a transverse (divergence-free and not rotation-free) field that should induce magnetic field~\cite{li1996jap,kline2024prapp}. This theoretical aspect will be discussed in a separate report~\cite{kawada2024arXiv}. 

In summary, we have studied the acoustic voltage in heavy metal (HM)/ferromagnet (FM) bilayers to identify the source of the spin current generated in the acoustic spin Hall effect. The magnetic field angle dependence of the acoustic voltage exhibits a phase shift that depends on the SAW propagation direction with respect to the crystalline orientation of the LiNbO$_3$ substrate. 
A model calculation shows that the results can be accounted for if an ac electric field propagating at the speed of the SAW, referred to as the acousto-electric evanescent wave, generates the spin current in the HM layer via the spin Hall effect.
These findings in turn imply potential applications of the acousto-electric evanescent waves. 
As their phase velocity is several orders of magnitude slower than that of light, the associated electric field has a wavenumber far larger than that of microwaves. 
In analogy with the excitation of surface plasmon polariton by evanescent light, the acousto-electric evanescent wave may provide a novel approach for probing unexplored quasiparticle excitations in materials that are inaccessible by conventional electromagnetic waves.

\section{Acknowledgments}
We thank K. Usami for fruitful discussions. This work was partly supported by JSPS KAKENHI (Grant Nos. 20J21915, 20J20952, 21K13886, 23KJ1419, 23KJ1159, 23H05463, and 24K00576), JST PRESTO Grant No. JPMHPR20LB, Japan, and JSPS Bilateral Program Number JPJSBP120245708.

\bibliography{reference_ASHE_main_arXiv}

\end{document}